\documentclass[a4paper,11pt]{article}
\usepackage{pos}

\title{The Fermi-LAT Light Curve Repository:\\ A resource for the time-domain and multi-messenger communities}
 \ShortTitle{LCR: A resource for the TDMM communities}

\author*[a,b]{Janeth Valverde}
\affiliation[a]{Department of Physics and Center for Space Sciences and Technology, University of Maryland Baltimore County, Baltimore, MD 21250, USA}
\affiliation[b]{NASA Goddard Space Flight Center, Greenbelt, MD 20771, USA}

\author[c]{D.~Kocevski}
\affiliation[c]{NASA Marshall Space Flight Center, Huntsville, AL 35812, USA}

\author[d]{M.~Negro}
\affiliation[d]{Department of Physics \& Astronomy, Louisiana State University, Baton Rouge, LA 70803, USA}

\author[e]{S.~Garrappa}
\affiliation[e]{Department of Particle Physics and Astrophysics, Weizmann Institute of Science, 76100 Rehovot, Israel}

\author[f]{A.~Brill}
\affiliation[f]{NASA Postdoctoral Program Fellow, USA}

\onbehalf{, for the Fermi-LAT collaboration} 

\emailAdd{janeth@umbc.edu}

\abstract{For over 15 years the Fermi Large Area Telescope (Fermi-LAT) has been monitoring the entire high-energy gamma-ray sky, providing the best sampled 0.1 -- $>1$ TeV photons to this day. As a result, the Fermi-LAT has been serving the time-domain and multi-messenger community as the main source of gamma-ray activity alerts. All of this makes the Fermi-LAT a key instrument towards understanding the underlying physics behind the most extreme objects in the universe. However, generating mission-long LAT light curves can be very computationally expensive. The Fermi-LAT light curve repository (LCR) tackles this issue. The LCR is a public library of gamma-ray light curves for 1525 Fermi-LAT sources deemed variable in the 4FGL-DR2 catalog. The repository consists of light curves on timescales of days, weeks, and months, generated through a full-likelihood unbinned analysis of the source and surrounding region, providing flux and photon index measurements for each time interval. Hosted at NASA's FSSC, the library provides users with access to this continually updated light curve data, further serving as a resource to the time-domain and multi-messenger communities. }

\ConferenceLogo{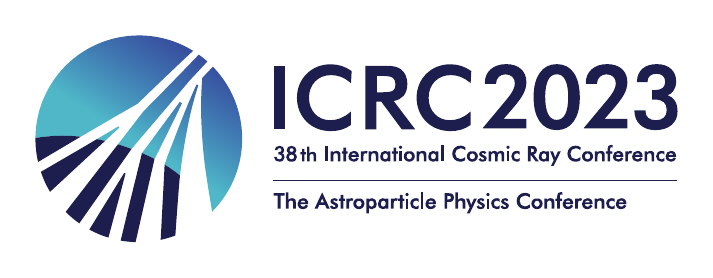}

\FullConference{%
38th International Cosmic Ray Conference (ICRC2023)\\
  26 July - 3 August, 2023\\
  Nagoya, Japan}

\begin{document}
\maketitle

\section{Introduction}

For over 15 years the {\sl Fermi} Large Area Telescope \citep[LAT;][]{theLAT} has enabled the identification and regular monitoring of over six thousand $\gamma$-ray transient, variable, and steady-state sources. 
The {\it Fermi}-LAT capabilities combined with a sky-scanning strategy for the great majority of the mission, allows it to reach an almost uniform full-sky coverage every two orbits, or approximately every three hours. This has made the LAT a pivotal tool in the study of time-domain and multi-messenger astronomy. For instance, the {\sl Fermi}-LAT played a crucial role in the possible first association of high-energy (HE) neutrinos detected by the IceCube neutrino observatory, and the flaring blazar TXS 0506+056 \citep{IceCube:2018dnn, IceCube:2018cha}. Such association could indicate the presence of ionized matter (i.e. cosmic rays, e.g. protons) in the emission region within the $\gamma$-ray blazar, making them a possible source of cosmic rays. Both, the nature of the HE emission from blazars, and the origin of cosmic rays have long been a subject of debate in the HE astrophysical community. 

Most {\sl Fermi}-LAT sources are also seen at other wavelengths, many times, in the whole electromagnetic spectrum. Long-term multi-wavelength and multi-messenger observations are warranted in order to probe models attempting to explain their broadband emission, in order to search for the correlated behavior that these models imply or predict, or in order to investigate their geometry (e.g. with quasi-periodic searches to probe the existence of binary systems or jet precession in active galactic nuclei) or accreting timescales of these objects (e.g. with temporal trends). The {\sl Fermi}-LAT has been indispensable to multi-wavelength campaigns that aim to perform these studies \citep[e.g.,][]{valverde20}. 

In the following, we will provide a brief overview of the {\sl Fermi}-LAT Light Curve Repository \citep[LCR;][]{lcr}\footnote{\url{https://fermi.gsfc.nasa.gov/ssc/data/access/lat/lcr/}}, the first publication-quality, continuously updated,  data products that the Fermi-LAT Collaboration has released. We will discuss the science that the community has exploited using the LCR data so far, and how it has been used to enable time-domain and multi-messenger (TDAMM) science.

\section{The LCR}

\begin{figure}[h]
\begin{center}
\includegraphics[width=.85\textwidth,angle=0]{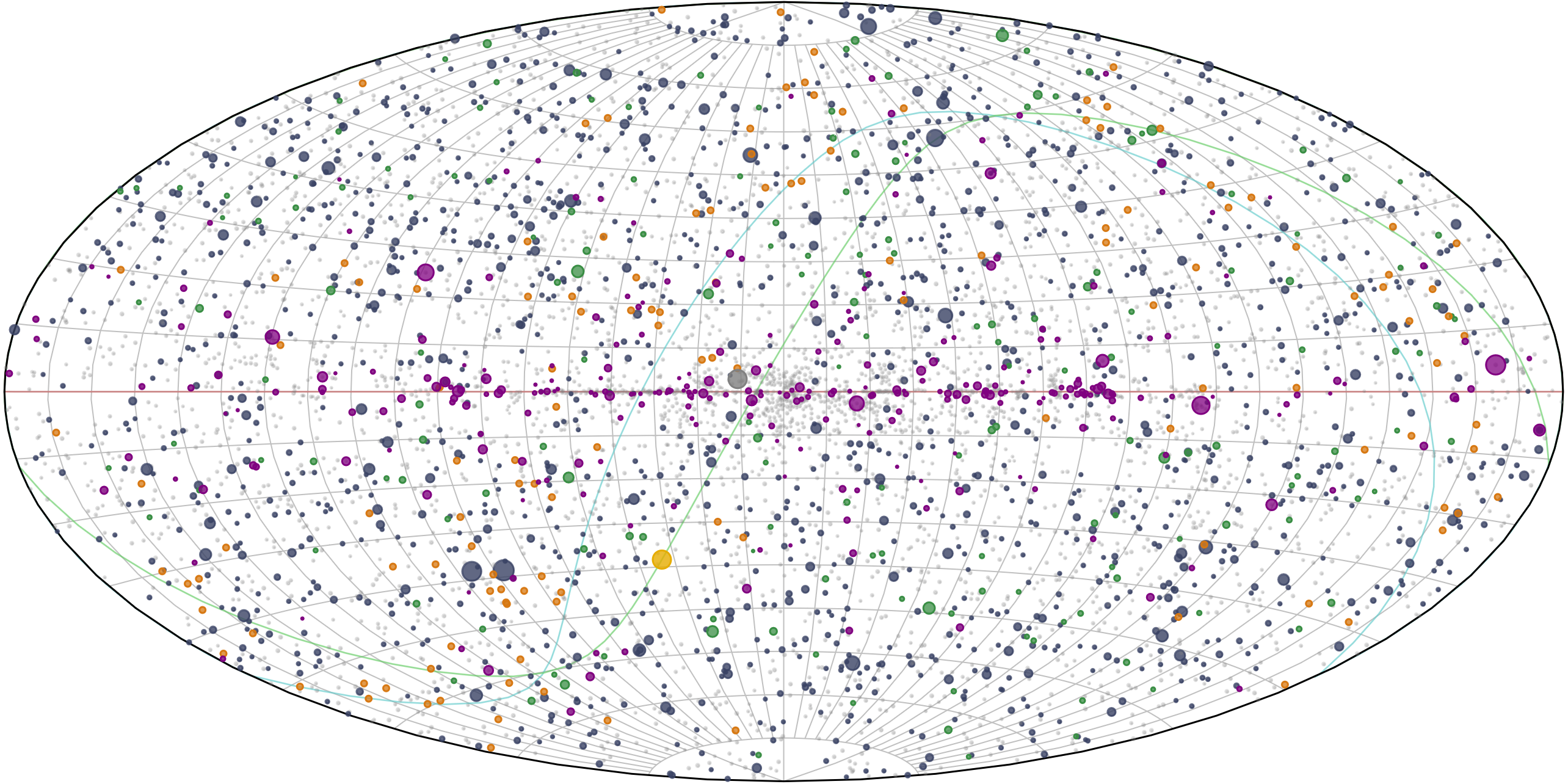}
\includegraphics[width=.85\textwidth,angle=0]{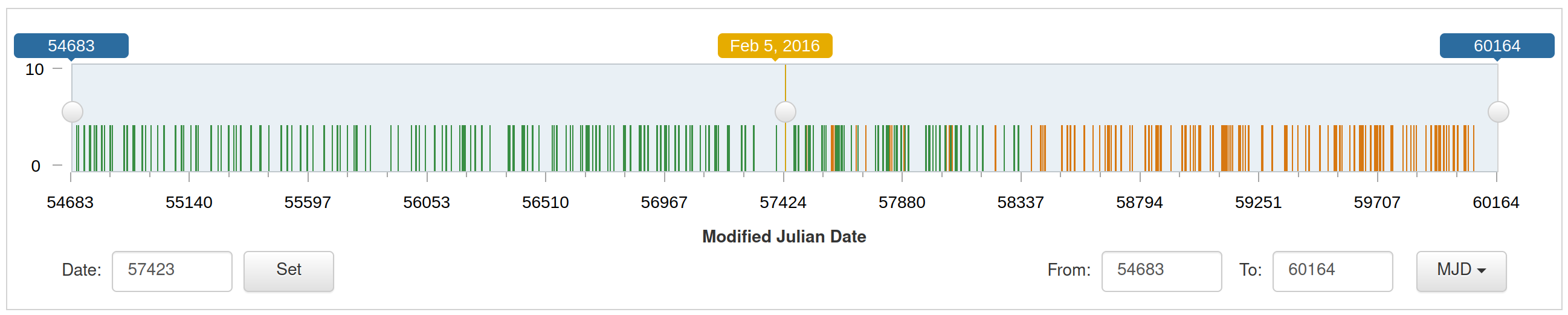}
\caption{Top: LCR main page map. Light gray markers show 4FGL-DR2 sources, while dark gray markers show LCR sources, i.e. 4FGL-DR2 sources with variability index higher than 21.67. 3PC sources  are shown in purple. Superimposed, in green, are the error circles from the IceCube neutrino alerts. The error circles from the second LAT GRB catalog are shown in orange. The ecliptic, equatorial and galactic planes, as well as the Sun and the Moon are also displayed. Double clicking on the markers will display a \textit{tooltip} box showing the source's key characteristics as well as linking to its LCR light curve, if applicable, its 4FGL-DR2 light curve and spectrum, among others, or its 3PC dedicated pages.
Bottom: If time-dependent data are plotted, such as IceCube or GRB positions or the marker size of the source position exhibiting time-resolved significance, an arrival-time plot is created beneath the map. 
\label{fig:map}}
\end{center}
\end{figure}

Obtaining whole-mission, publication quality, {\sl Fermi}-LAT light curves requires a full likelihood analysis of the region that takes into account the flux variations of all nearby sources. However, generating a high cadence (e.g., in daily timescales) light curve using a full likelihood treatment over the entire duration of the mission can be very computationally expensive. 

The LCR tackles this issue by distributing the analyses of the light curve bins to separate nodes in a computer cluster hosted at the SLAC National Accelerator Laboratory.
It is hosted at NASA's Fermi Science Support Center (FSSC), and consists of a public database of publication-quality light curves for variable {\sl Fermi}-LAT sources on three timescales, 3 days, 7 days (weekly), and 30 days (monthly), for 1525 sources deemed variable in the 4FGL-DR2 catalog \citep{4fgl-dr2}. The repository provides light curves generated through full likelihood analyses of the sources and surrounding regions, providing flux and spectral index measurements for each time bin, and it is continuously updated as new data becomes available. Two forms of light curves are provided for each cadence and for each source, one with the spectral index fixed, and another one with spectral index free, so that the user can cross check these light curves for consistent results. The LCR was promised in 2018 Senior Review as a new resource to the time-domain and multi-messenger communities for associating and monitoring LAT sources, in particular facilitating the identification of time intervals with high fluxes; and has been supported by Fermi Guest Investigator program\footnote{\url{https://fermi.gsfc.nasa.gov/ssc/proposals/}}. 

The LCR uses unbinned maximum likelihood analysis, in which the parameters of a model describing the point sources and diffuse isotropic $\gamma$-ray emission in a given region of the sky are jointly optimized to best describe the observed photon distribution. For each source and time bin, photons are selected from a circular region of interest (ROI) of radius 12$^{\circ}$ centered on the location of the target source. For a detailed review of the LCR analysis details, please refer to \cite{lcr}. Motivated by the science described above, the LCR provides light curves on the sources in the 4FGL-DR2 catalog that have variability indices greater than 21.67. Sources with such a variability index over 10 years are estimated to have a less than 1\% chance of being steady, based on the yearly light curves that the 4FGL-DR2 provides for the first 10 years of the mission. The result is a sample of 1525 sources, or 26\% of the 4FGL-DR2 catalog. The vast majority of these sources are blazars, making up about 93\% of the repository sample. 

The LCR data can be downloaded using an Application Programming Interface, (or API), links that can be integrated into scripts for automated analyses. The LCR website also provides an example of such a script\footnote{\url{https://fermi.gsfc.nasa.gov/ssc/data/access/lat/lcr/download.py}}. Daniel Kocevski, the Principal Investigator of the LCR project, has also made available pyLCR\footnote{\url{https://github.com/dankocevski/pyLCR/tree/main}}, which is a python interface for the LCR. The data is available in two formats, CSV and JSON. 

As described in the LCR paper \citep{lcr}, because the data is published as soon as it is available, the results are not validated by the LAT Collaboration prior to release. It is warranted that the users perform sanity checks by examining the ratio of flux to flux uncertainty vs. the square root of the TS, the distributions of fit results, e.g., ﬂux, ﬂux uncertainties, spectral indices mean values. A small percentage of the analyses might not converge (fit quality different from zero) or provide negative TS values for the targets; these results are not reliable and should not be used. Some examples of these are shown in the ﬂux over the ﬂux uncertainty ratio is expected to be approximately proportional to the square root of the TS. Any outliers should be either further investigated or removed before using the data for higher-level analyses, as should any outliers from the data distributions. Details on how to perform these sanity checks can be found both in \cite{lcr} and in Chapter 4 of \cite{thesis20}. 
In \citep{lcr}, we also elaborate on other caveats that user should be aware of while using these data. All of these are caveats that can be present in any {\sl Fermi}-LAT likelihood analysis, so the user should be aware of them even if they perform their own LAT fits and do not use the LCR data. 

The main page of the LCR website features an interactive map that display the positions of all 4FGL-DR2 sources. The user can select what the marker size in this map will indicate: 4FGL-DR2 Variability index, 4FGL-DR2 average Significance, or LCR 3-day time-resolved significance. 
Optionally, additional data may be overlaid on the map: Real time Sun or Moon position, positions of IceCube neutrino alerts, gamma-ray burst (GRB) error circles as reported in the Second LAT GRB catalog \citep{grb2}, and, very recently added, the third LAT Pulsar Catalog \citep[3PC;][]{2023arXiv230711132S}\footnote{\url{https://fermi.gsfc.nasa.gov/ssc/data/access/lat/3rd_PSR_catalog/}} that contains data products dedicated to pulsar science on 294 pulsars. The 1525 LCR sources are highlighted in dark gray, while by default the non-variable 4FGL-DR2 sources are marked in light gray. See Fig. \ref{fig:map}. Hovering over any source displays a \textit{tooltip} box showing its name and key characteristics as well as linking to its 4FGL light curve and spectrum, related 3PC entry, related Fermi All-sky Variability Analysis \citep[FAVA;][]{2017ApJ...846...34A} entry, and LCR light curve if applicable.

\section{Scientific impact}

Since its public release in December 2021, the community has been using the LCR data to exploit, so far, a number of active galactic nuclei science topics. A recent study \citep{brill22} used deep learning techniques to investigate the structure in gamma-ray blazars’ complex variability patterns that traditional analysis methods may miss. They applied self-supervised deep learning to create a model that encodes a representation of blazar variability. Their model predicts the flux probability distribution at each time step. Their model output can be analyzed to extract scientifically relevant information, like changes in weekly-timescale flux distributions over time or between sources. 

Other two studies \citep{2023MNRAS.518.5788O, 2023ApJ...950..173D} searched for quasi-periodic oscillations in a sample of bright LCR blazars. The former \citep{2023MNRAS.518.5788O} used the LCR and multi-wavelength data in order to probe the presence of a binary supermassive black hole system, or geometrical effects like helical or precessing jets. The latter \citep{2023ApJ...950..173D} used LCR data to argue that the current-driven kink instability and curved jet model could be the most likely causes for shorter and longer QPOs. Neither found conclusive evidence for the presence of QPOs in the sources they studied. In order to probe leptonic and hadronic models, another study \citep{2023MNRAS.519.6349D} used optical and LCR light curve data to search for orphan and non-orphan flares, and, in the case of the latter, to determine time-lags between the flare detection in the two energy bands. 

Two production channels were used \cite{2022PhRvD.106l3005S}, electron synchrotron self Compton and proton synchrotron, in order to model the combined HE ($0.1<E<100$ GeV) with very HE ($>0.1$ TeV) gamma-ray emission from the neutrino blazar TXS 0506 +056, detected 45 days after an LCR flare. Optical and LCR light curves have also been used \cite{2022MNRAS.516.2671P} to investigate their temporal variability and correlated patterns, and separate the non-thermal from the thermal contributions. They used this information, and a reverberation mapping technique, in order to estimate the size of the broad-line region and the black hole mass of the flat spectrum radio quasar (FSRQ) PKS 0736$+$017.

\subsection{The LCR in public alerts to the multi-wavelength and multi-messenger communities}

\begin{figure}[h]
\begin{center}
\includegraphics[width=.32\textwidth,angle=0]{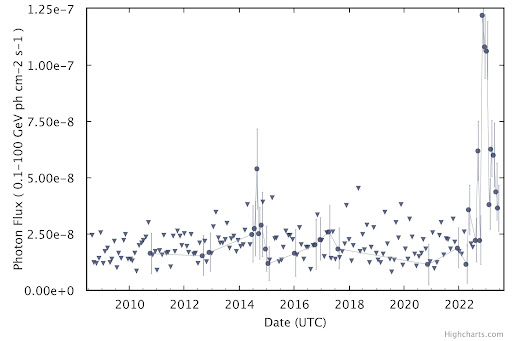}\hspace{0.01\textwidth}
\includegraphics[width=.32\textwidth,angle=0]{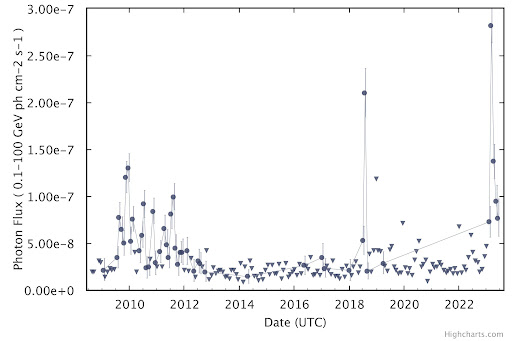}\hspace{0.01\textwidth}
\includegraphics[width=.32\textwidth,angle=0]{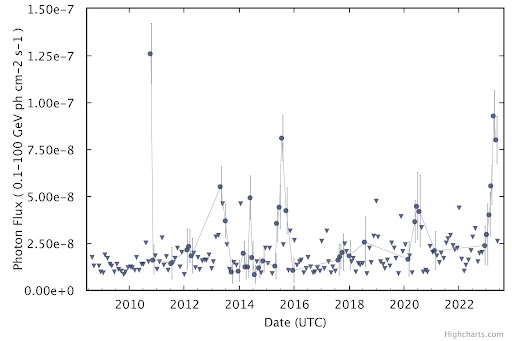}
\includegraphics[width=.32\textwidth,angle=0]{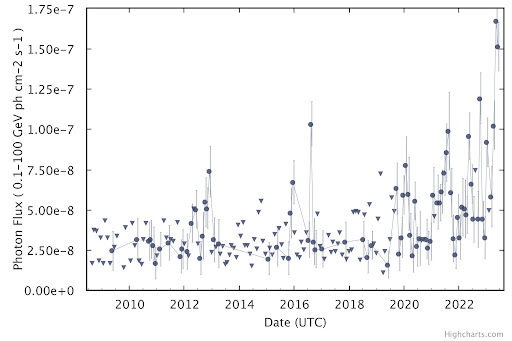}\hspace{0.01\textwidth}
\includegraphics[width=.32\textwidth,angle=0]{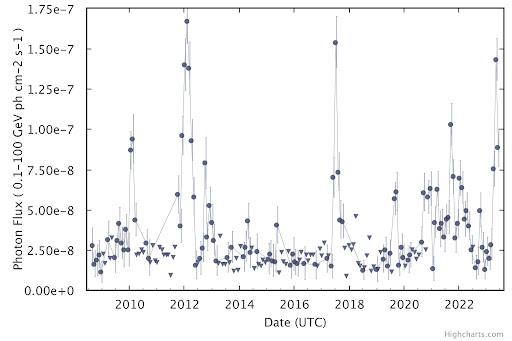}\hspace{0.01\textwidth}
\includegraphics[width=.32\textwidth,angle=0]{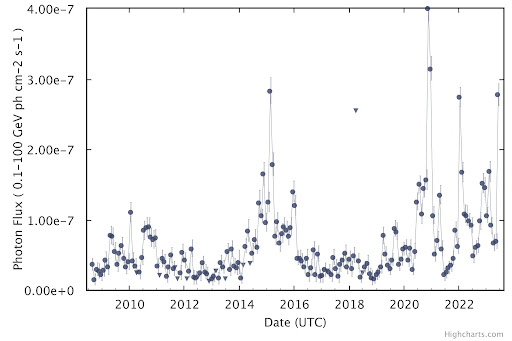}
\includegraphics[width=.32\textwidth,angle=0]{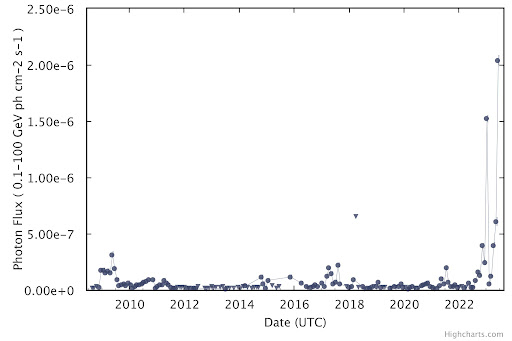}\hspace{0.01\textwidth}
\includegraphics[width=.32\textwidth,angle=0]{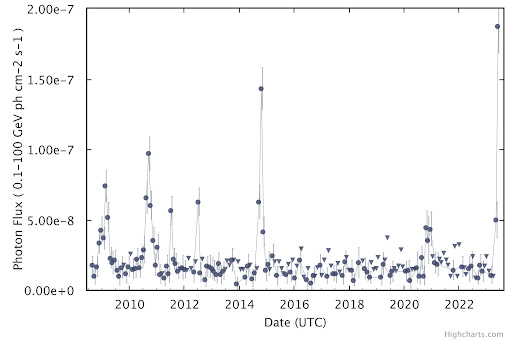}\hspace{0.01\textwidth}
\includegraphics[width=.32\textwidth,angle=0]{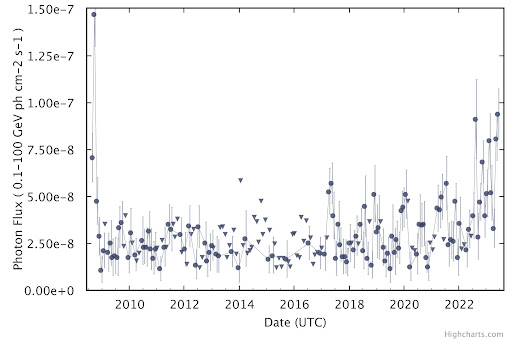}
\caption{LCR light curves. It spans more than 15 years of LAT data from some of the sources that showed relatively high $\gamma$ activity. From top to bottom and left to right: FSRQ TXS 0007$+$205 (with redshift $z=0.597$), blazar PMN J1830-4441 ($z$ unknown), FSRQ PMN J1913-3630 ($z$ unknown), FSRQ PKS 2245-328($z=2.268$), FSRQ PKS 0700-465 ($z=0.822$), FSRQ S4 1726$+$45 ($z=0.717$), FSRQ 4C $+$31.03 ($z=0.603$), BL Lac object MG2 J043337$+$2905 ($z=0.97$), FSRQ PKS 2204-54 ($z=1.206$). From \url{https://fermi.gsfc.nasa.gov/ssc/data/access/lat/lcr/}.
\label{fig:lcs}}
\end{center}
\end{figure}

The scientific exploitation of the {\sl Fermi}-LAT capabilities and observational strategy is enhanced when early information about transients reaches a broader astrophysical community. The {\sl Fermi}-LAT Collaboration put in place a Gamma-ray Sky Watcher program, since almost the start of the mission, run by the Flare Advocate (FA) group. The FAs work on weekly shifts to validate results and quickly issue public alerts about flares and new detections, thus triggering multi-wavelength and multi-messenger follow-ups \citep{2012AIPC.1505..697C}. Therefore, these activities are key to scientific results. 

Since its public release on December 2021, the LCR has been an extremely valuable tool for the FAs, allowing them to cross check flaring activities and their timescales. At the time of writing, the LCR has been cited in over 53 {\sl Fermi}-LAT Astronomer's Telegrams and Gamma-ray Circular Notices\footnote{\url{https://www-glast.stanford.edu/cgi-bin/pub_rapid}} broadcasting renewed or enhanced flaring activity of associated {\sl Fermi}-LAT sources, or in response to neutrino alerts when LCR sources happen to be within or close to the neutrino alert error circle. Fig. \ref{fig:lcs} shows the LCR light curves of some of the sources that exhibited a flaring episode within the last six months. 

\section{Summary}

For the first time, the {\sl Fermi}-LAT Collaboration has released publication-quality, continuously updated,  data products, the LCR. The science themes and studies it has started to enable are only growing in number, and is proving to be an invaluable resource for the time-domain and multi-messenger communities.

\section{Acknowledgments}
This material is based upon work supported by NASA under award number 80GSFC21M0002. 
The \textit{Fermi}-LAT Collaboration acknowledges support for LAT development, operation and data analysis from NASA and DOE (United States), CEA/Irfu and IN2P3/CNRS (France), ASI and INFN (Italy), MEXT, KEK, and JAXA (Japan), and the K.A.~Wallenberg Foundation, the Swedish Research Council and the National Space Board (Sweden). Science analysis support in the operations phase from INAF (Italy) and CNES (France) is also gratefully acknowledged. This work performed in part under DOE Contract DE-AC02-76SF00515.

%
%
%

\end{document}